\documentclass[12pt]{iopart}
\pdfoutput=1
\usepackage{iopams}
\usepackage{geometry}
\usepackage{graphicx}
\usepackage{multirow}

\begin{document}

\title[Monte Carlo simulation of MLC-shaped TrueBeam electron fields]{Monte Carlo simulation of MLC-shaped TrueBeam electron fields benchmarked against measurement}
\author{Samantha AM Lloyd$^1$, Isabelle M Gagne$^{2, 1}$ and Sergei Zavgorodni$^{2, 1}$\\
		$^1$ Department of Physics and Astronomy, University of Victoria, Victoria, BC, Canada\\
		$^2$Department of Medical Physics, BC Cancer Agency - Vancouver Island Centre, Victoria, BC, Canada}

\ead{slloyd@bccancer.bc.ca}

%%% 	BEGIN DOCUMENT		%%%

\begin{abstract}
%\begin{spacing}{1.3}

	Modulated electron radiotherapy (MERT) and combined, modulated photon and electron radiotherapy (MPERT) have received increased research attention, having shown capacity for reduced low dose exposure to healthy tissue and comparable, if not improved, target coverage for a number of treatment sites. Accurate dose calculation tools are necessary for clinical treatment planning, and Monte Carlo is the gold standard for electron field simulation. With many clinics replacing older accelerators with newer machines, Monte Carlo source models of these accelerators are needed for continued development, however, Varian has kept internal schematics of the TrueBeam confidential and electron phase-space sources have not been made available. Because TrueBeam electron fields are not substantially different from those generated by the Clinac 21EX, we have modified the internal schematics of the Clinac 21EX to simulate TrueBeam electrons with MLC-shaped apertures. BEAMnrc and DOSXYZnrc were used to simulate $5\times5$ and $20\times20$ cm$^2$ electron fields with MLC-shaped apertures at 6, 12 and 20 MeV. Secondary collimating jaws were set 0.5 cm beyond the MLC periphery and to $40\times40$ cm$^2$. Our complete accelerator models agreed with diode measurement within 2\%/2 mm at 12 and 20 MeV, and, for the most part, within 3\%/3 mm at 6 MeV. Comparisons of measured depth and profile data showed dose-dependencies on jaw position; simulated charged particle energy fluences scored just above the phantom showed that, for small apertures, dose dependencies on jaw position are dominated by changes in the in-field energy fluence (not scattered by the jaws or MLC) at 6 MeV, while at 20 MeV dose dependencies are dominated by the scattered component. Our models reproduce these jaw position dependencies, which is an asset as there is no consensus on the optimal position for jaws in modulated electron field delivery. Good agreement between simulation and measurement and flexibility in jaw position make our Monte Carlo models appropriate for use in MERT and MPERT planning.

%\end{spacing}	
\end{abstract}


\maketitle

%%%	 	INTRODUCTION		%%%
%\begin{spacing}{1.4}
\section{Introduction}

	Modulated electron radiotherapy (MERT) and combined, modulated photon and electron radiotherapy (MPERT) have been shown, through retrospective planning studies, to reduce dose delivered to healthy tissue and/or to improve target dose uniformity for some breast \cite{Alexander:2011, Xiong:2004, Ma:2003}, post-mastectomy chest wall \cite{Gauer:2010, Salguero:2009}, head and neck \cite{Salguero:2010} and scalp treatments \cite{Jin:2008} compared to conventional electron therapies, and photon-IMRT. Although MERT and MPERT have seen limited clinical application, in-house planning and delivery systems are being designed and validated in efforts toward broader clinical utility \cite{Engel:2009, Alexander:2010}.

	The advancement of electron treatments involving complex modulation relies on the accuracy of the planning tools utilized. Monte Carlo techniques are well recognized as the gold standard for radiation transport simulation \cite{Reynaert:2007} and, due to the challenges associated with simulating irregularly shaped electron fields, are generally preferred for MERT and MPERT planning. A common challenge in the application of Monte Carlo techniques for electron field modelling is the development of accurate source models. One approach, presented by Ma et al. \cite{Ma:1997}, discussed the development and use of multi-source models in place of full accelerator models. This approach was implemented by Jiang et al. \cite{Jiang:2000} resulting in a four-source model for a Varian (Varian Medical Systems, Palo Alto, CA) Clinac 2100C that could be tuned and applied to multiple machines of the same design and agreed with complete accelerator simulations within 2\%/2 mm. Papaconstadopoulus and Seuntjens \cite{Papaconstadopoulos:2013} presented a Clinac 21EX source model that separated the beam into a primary source and multiple scattered sources, which agreed with full simulations within 3\%. Most recently, Henzen et al. \cite{Henzen:2014} presented a dual electron source model for Varian TrueBeam and Clinac 23EX linear accelerators that utilized a primary foil source and secondary jaw source for fixed jaw settings, which agreed with measurement within 3\%/3 mm.
	
	While multi-source models have advantageous calculation times, their efficiencies are bought through approximations. Furthermore, the accuracy of these approaches is generally measured against complete accelerator simulations. Complete models explicitly simulate the geometry of the linear accelerator head and provide a tool for exploring the characteristics of particle fluence at any location within the model. Klein et al. \cite{Klein:2008}, Al-Yahya et al. \cite{AlYahya:2005mp}, and Salguero et al. \cite{Salguero:2009} used complete accelerator models to simulate Varian Trilogy, Clinac 2100EX, and Siemens (Siemens Healthcare, Erlangen, Germany) Primus electron sources, respectively, with overall agreements within 3\%/3 mm compared to measurement.
	
	The study by Klein et al. and another by du Plessis et al. \cite{duPlessis:2006} characterized the use of the photon multi-leaf collimators (MLCs) inherent on Trilogy and Primus accelerators to form electron field apertures. In order to mitigate the degradation of penumbra definition due to electron scatter, the source to surface distances (SSDs) used were reduced to between 60 and 85 cm. The Klein et al. study found the Varian MLC to be appropriate for electron field modulation at 70 and 85 cm SSD, with limitations set on the smallest available field size as a function of SSD, while du Plessis et al. found the Siemens MLC to maintain field definition up to 70 cm SSD for low-energy electrons, with the least penumbra broadening at 60 cm SSD for all energies. Following du Plessis's characterization, Salguero et al. showed the Primus accelerator's MLC-shaped electron apertures to be appropriate for MERT delivery, achieving clinically appropriate coverage and uniformity that was, in some cases, dosimetrically advantageous compared to the current clinical standard (photon-IMRT or 3D-conformal radiotherapy) for post-mastectomized chest walls \cite{Salguero:2009} and shallow head and neck tumours \cite{Salguero:2010}. 
	
	Other groups have approached the issue of penumbra degradation by designing and using tertiary electron-specific multi-leaf collimators (eMLCs) positioned close to the patient surface. The Clinac 2100EX accelerator model presented by Al-Yahya et al. included a few-leaf electron collimator (FLEC), the design and testing of which was presented in an earlier paper \cite{AlYahya:2005pmb}. Gauer et al. \cite{Gauer:2008} and Jin et al. \cite{Jin:2013} presented and characterized prototype eMLC designs while Jin et al. also presented a complementary Monte Carlo model of the device.  Although these in-house systems produced clinically appropriate electron fields at standard delivery distances, photon-MLCs are inherent in most clinical accelerators and Monte Carlo models are available for the most common designs, including the Varian Millennium MLC available on many Varian machines.

	 Due, in part, to the proprietary nature of its internal schematics, Varian has not released Monte Carlo specifications or phase-space sources for TrueBeam electron fields. Given that the characteristics of electron fields generated by a TrueBeam accelerator do not deviate significantly from the characteristics of electron fields generated by a Clinac 21EX accelerator, and that the internal schematics of a Clinac 21EX accelerator are known, a complete Monte Carlo accelerator model can be built to match the output of a TrueBeam accelerator by modifying a Clinac 21EX accelerator model. The aim of this work is to present such models for 6, 12 and 20 MeV TrueBeam electron sources. The robustness of these modified models was tested against measurements of MLC-shaped electron fields at two aperture sizes. To investigate the role of the secondary collimating jaws in electron field shaping, each aperture was examined at two representative jaw settings: close to the aperture periphery, and completely retracted. The accuracy of these models is essential in the development of Monte Carlo-based MERT and MPERT planning and optimization systems for TrueBeam deliveries.

%%%		METHODS		%%%

\section{Methods}

\subsection{Diode Measurements}
\label{subsec:diodeMeas}

	Measurements of TrueBeam electron field depth dose curves and dose profiles at nominal d$_\mathrm{max}$ and d$_{50}$ were performed using an EFD$^\mathrm{3G}$ scanning electron field diode with a CC13 ion chamber reference in a large water tank ($48\times48\times41$ cm$^3$). Measurements were acquired with OmniPro v7.4 (IBA Dosimetry, Schwarzenbruck, Germany) for 6, 12, and 20 MeV electron fields at 70 cm SSD. For all energies, both $5\times5$ and $20\times20$ cm$^2$ MLC-shaped apertures were delivered, and for each aperture size, two jaw settings were used. The first jaw setting was $40\times40$ cm$^2$ for both MLC aperture sizes, while the second settings placed the jaws at $6\times6$ and $21\times21$ cm$^2$ for the $5\times5$ and $20\times20$ cm$^2$ MLC apertures, respectively (jaws set to 0.5 cm beyond the MLC periphery). All jaw and aperture sizes are specified by their projected size at isocenter and nominal values of d$_\mathrm{max}$ and d$_{50}$ were such that d$_\mathrm{max} = 1.4$, 2.7, 2.2 cm and d$_{50} = 2.4$, 5.1, 8.3 cm for 6, 12 and 20 MeV, respectively. Profiles were measured in the crossline and inline directions where crossline is perpendicular to the direction of the waveguide and inline is parallel to the waveguide.
	
	Measured data was corrected for noise by the reference, smoothed and centred in OmniPro, then exported for processing in MATLAB (MathWorks, Natick, MA). To preserve and compare relative differences in output, measurements were scaled by a select value for each aperture-energy combination; this value was chosen so that data acquired with jaws set to $40\times40$ cm$^2$ were normalized to 100\% at nominal d$_\mathrm{max}$ for all aperture sizes and energies.

\subsection{Monte Carlo Simulations}

	Monte Carlo simulations were performed using the Vancouver Island Monte Carlo System \cite{Zavgorodni:2007, Bush:2008}, which uses the EGSnrc-based \cite{Kawrakow:2006} user codes BEAMnrc \cite{Rogers:2009} and DOSXYZnrc \cite{Walters:2011}. Photon and electron cut-off energies were set to AP = PCUT = 0.01 MeV and AE = ECUT = 0.7 MeV, respectively. Dose was calculated in a $30\times30\times30$ cm$^2$ water phantom with a $0.2\times0.2\times0.2$ cm$^3$ voxel grid. In addition to dose distributions, phase-space files were scored above the phantom to analyze the contribution of jaw- and MLC-scattered electrons to energy fluence and dose. The LATCH option in BEAMnrc was used to track particles that interacted or originated in the jaws or MLC during the simulation (LATCH option 3). The program BEAMDP \cite{Ma:2004} was used to extract energy fluence information from resulting phase-spaces.

\begin{figure}[h]
  \caption{Schematic of component modules used in the complete BEAMnrc/DOSXYZnrc accelerator models of TrueBeam electron sources.}
  \centering
  \includegraphics[width=0.60\textwidth]{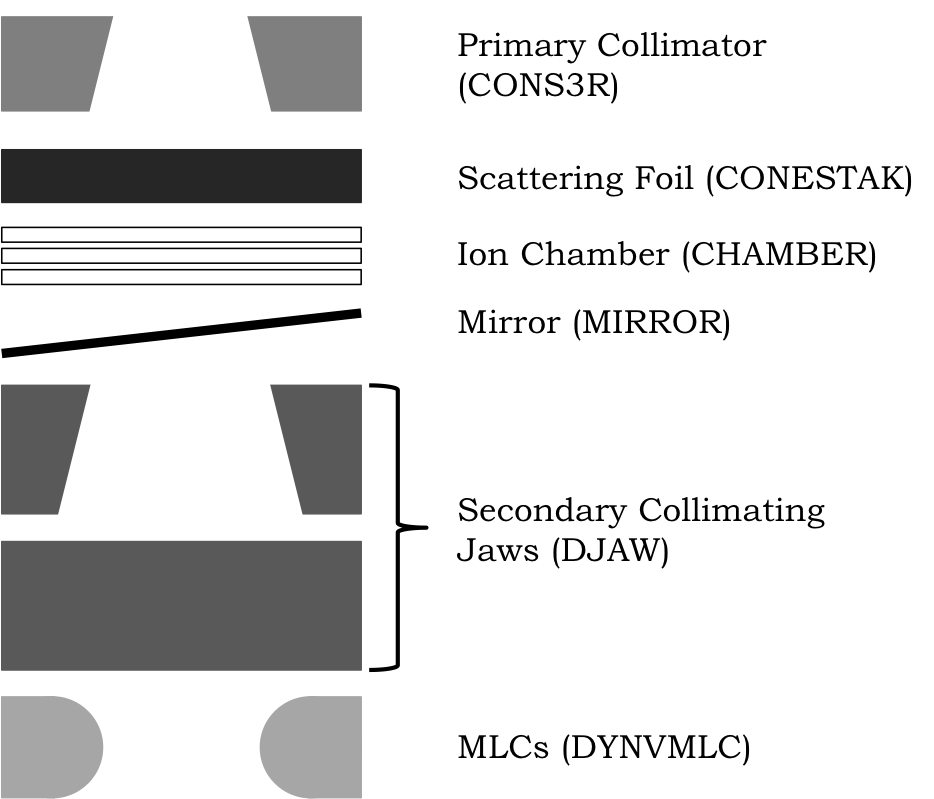}
  \label{fig:accelerator}
\end{figure}

	As internal schematics have not been made available, the accelerator models used for TrueBeam simulations are based on the manufacturer specified schematics of a Clinac 21EX accelerator with modifications made to the scattering foils. Each model uses a forwarded directed source of mono-energetic electrons with a circular, Gaussian spatial distribution (ISOURC = 19). The incident electron energy and FWHM of the Gaussian spread for each model are specified in Table \ref{tab:TBsource}. The component modules used to build the TrueBeam model are listed in Figure \ref{fig:accelerator}.

\begin{table}[h]
  \centering
  \begin{tabular}{ c  c  c }
  \hline \hline
  Nominal Energy	& Incident Energy	& Gaussian FWHM \\
  (MeV)			& (MeV)			& (cm)\\
  \hline
  6				& 6.75			& 0.65\\
  12				& 13.38			& 0.45\\
  20				& 22.10			& 0.35\\
  \hline \hline
  \end{tabular}
  \caption{Source parameters for incident electrons in TrueBeam electron Monte Carlo models.}
  \label{tab:TBsource}
\end{table}

%%%		RESULTS			%%%

\section{Results}

\subsection{TrueBeam measurements and jaw settings}

\begin{figure}[h]
  \caption{Relative depth dose and crossline profile measurements of TrueBeam electrons at two jaw settings for each MLC-shaped aperture demonstrating differences in output. $5\times5$ cm$^2$ aperture measurements are shown on the left while $20\times20$ cm$^2$ aperture measurements are shown on the right. Data for jaws set to $40\times40$ cm$^2$ is plotted in solid lines while data for jaws set to $6\times6$ cm$^2$ (left) and $21\times21$ cm$^2$ (right) is plotted in dashed lines.}
  \centering
  \includegraphics[width=1.0\textwidth]{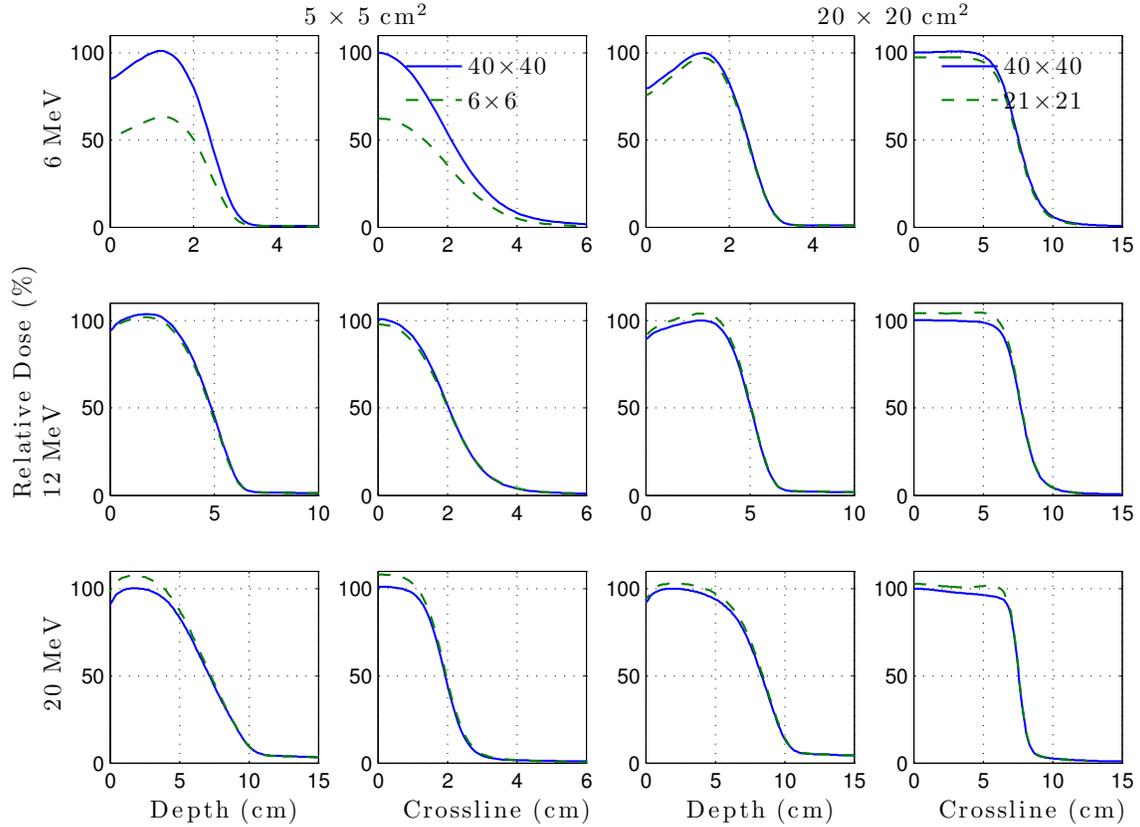}
  \label{fig:jawMeas}
\end{figure}

	Figure \ref{fig:jawMeas} shows diode measurements of electron fields from a TrueBeam accelerator for $5\times5$ and $20\times20$ cm$^2$ MLC-shaped apertures with jaws set to $6\times6$ and $21\times21$ cm$^2$, respectively, as well as to $40\times40$ cm$^2$ as specified in Section \ref{subsec:diodeMeas}. The dose magnitude and characteristic dependencies on jaw position vary with both aperture size and electron energy. Of note is the 35\% decrease in dose magnitude for a $5\times5$ cm$^2$ aperture of 6 MeV electrons when jaws are moved in from $40\times40$ to $6\times6$ cm$^2$.

	As expected, crossline data at all energies show no discernible difference in out-of-field dose between jaw settings, indicating that electron transmission through the MLC is negligible when the jaws are retracted. This means that, for MERT treatments, jaws are not required in order to minimize MLC transmission dose.

\subsection{TrueBeam Monte Carlo model}

\begin{figure}[ht!]
  \caption{Relative depth and dose profile diode measurements plotted against Monte Carlo (histogram) along the central beam axis for $5\times5$ and $20\times20$ cm$^2$ apertures with jaws set to $6\times6$ and $21\times21$ cm$^2$. Data is shown for 6 MeV (left), 12 MeV (centre) and 20 MeV (right)}
  \centering
  \includegraphics[width=0.70\textwidth]{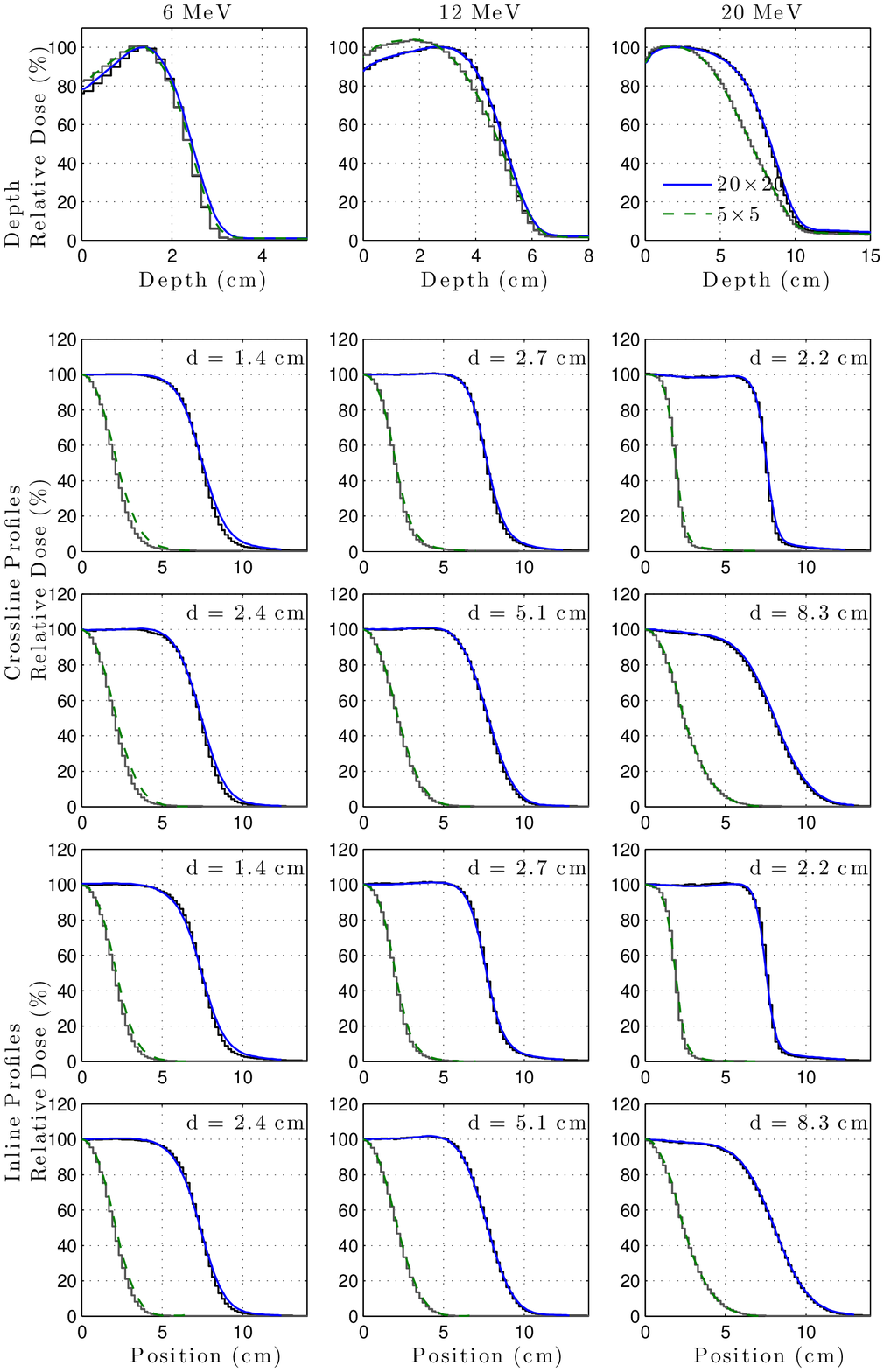}
  \label{fig:jawInMC}
\end{figure}

\begin{figure}[ht!]
  \caption{Relative depth and dose profile diode measurements plotted against Monte Carlo (histogram) along the central beam axis for $5\times5$ and $20\times20$ cm$^2$ apertures with jaws set to $40\times40$ cm$^2$. Data is shown for 6 MeV (left), 12 MeV (centre) and 20 MeV (right).}
  \centering
  \includegraphics[width=0.7\textwidth]{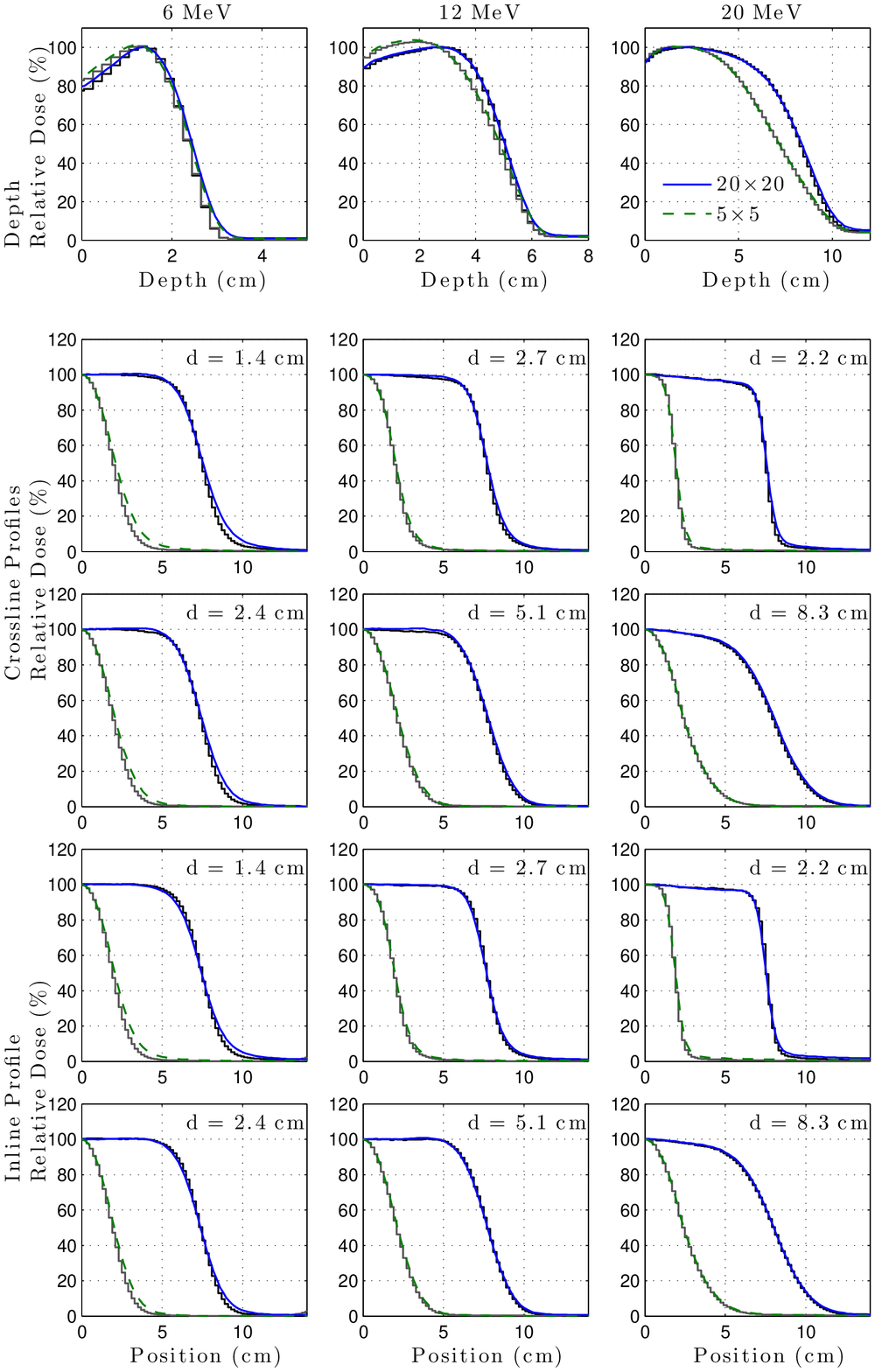}
  \label{fig:jawOutMC}
\end{figure}

	Figure \ref{fig:jawInMC} shows Monte Carlo data plotted against measurements for $5\times5$ and $20\times20$ cm$^2$ apertures with jaws set to $6\times6$ and $21\times21$ cm$^2$, respectively. Figure \ref{fig:jawOutMC} shows Monte Carlo data plotted against measurements for the same fields, but with jaws set to $40\times40$ cm$^2$ in all cases. Profiles are plotted in both the crossline and inline orientations at depths of nominal d$_\mathrm{max}$ and d$_{50}$. Depth dose data is normalized to 100\% at nominal d$_\mathrm{max}$ while all profile data has been normalized to 100\% along the central axis.

\begin{table}[ht]
  \centering
  \begin{tabular}{ c  c  c  c  c  c  }
  \hline
  Jaw Position		& \multicolumn{2}{c}{$6\times6$ cm$^2$}		&	& \multicolumn{2}{c}{$40\times40$ cm$^2$} \\
  							& Diode	& Monte Carlo		&	& Diode	& Monte Carlo \\
  \hline \hline
  							& \multicolumn{5}{c}{6 MeV} \\
  %d$_\mathrm{max}$ (cm)			& 1.3		& 1.3 			&	& 1.2		& 1.3 \\
  d$_{90}$	 (cm)					& 1.9		& 1.9 			&	& 1.9		& 1.9 \\
  d$_{50}$	 (cm)					& 2.4		& 2.4 			&	& 2.4		& 2.4 \\
  d$_{20}$	 (cm)					& 2.8		& 2.7 			&	& 2.8		& 2.7 \\
  FWHM$_\mathrm{cross}$(cm)	& 4.3		& 4.0 			&	& 4.2		& 4.0 \\
  FWHM$_\mathrm{in}$ (cm)		& 4.4		& 4.1 			&	& 4.1		& 3.8 \\
  80-20\%$_\mathrm{cross}$ (cm)	& 1.9		& 1.6	 			&	& 1.9		& 1.6 \\
  80-20\%$_\mathrm{in}$ (cm)		& 1.6		& 1.5	 			&	& 1.8		& 1.5 \\
							& \multicolumn{5}{c}{12 MeV} \\
  %d$_\mathrm{max}$ (cm)			& 1.7		& 1.8 			&	& 1.8		& 1.9 \\
  d$_{90}$	 (cm)					& 3.3		& 3.3 			&	& 3.2		& 3.3 \\
  d$_{50}$	 (cm)					& 4.8		& 4.7 			&	& 4.8		& 4.8 \\
  d$_{20}$	 (cm)					& 5.6		& 5.6 			&	& 5.7		& 5.6 \\
  FWHM$_\mathrm{cross}$(cm)	& 4.1		& 4.0 			&	& 4.1		& 4.0 \\
  FWHM$_\mathrm{in}$ (cm)		& 4.0		& 3.9 			&	& 3.9		& 3.9 \\
  80-20\%$_\mathrm{cross}$ (cm)	& 1.5		& 1.3	 			&	& 1.4		& 1.3 \\
  80-20\%$_\mathrm{in}$ (cm)		& 1.4		& 1.4	 			&	& 1.4		& 1.3 \\
    							& \multicolumn{5}{c}{20 MeV} \\
  %d$_\mathrm{max}$ (cm)			& 1.6		& 1.9 			&	& 1.8		& 1.6 \\
  d$_{90}$	 (cm)					& 4.3		& 4.3 			&	& 4.4		& 4.4 \\
  d$_{50}$	 (cm)					& 7.1		& 7.0 			&	& 7.2		& 7.1 \\
  d$_{20}$	 (cm)					& 9.0		& 9.0 			&	& 9.1		& 9.0 \\
  FWHM$_\mathrm{cross}$(cm)	& 3.8		& 3.8 			&	& 3.9		& 3.8 \\
  FWHM$_\mathrm{in}$ (cm)		& 3.7		& 3.8 			&	& 3.8		& 3.8 \\
  80-20\%$_\mathrm{cross}$ (cm)	& 0.9		& 0.8 			&	& 0.8		& 0.7 \\
  80-20\%$_\mathrm{in}$ (cm)		& 0.9		& 0.8	 			&	& 0.9		& 0.8 \\
  \hline \hline
  
  \end{tabular}
  \caption{Measured and simulated dose parameters for TrueBeam electron fields at 6, 12 and 20 MeV for $5\times5$ cm$^2$ apertures with jaws set to $6\times6$ and $40\times40$ cm$^2$. Width metrics are given in crossline (cross) and inline (in) orientations.}
  \label{tab:doseChar05}
\end{table}

\begin{table}[ht]
  \centering
  \begin{tabular}{ c  c  c  c  c  c  }
  \hline
  Jaw Position		& \multicolumn{2}{c}{$21\times21$ cm$^2$}	&	& \multicolumn{2}{c}{$40\times40$ cm$^2$} \\
  							& Diode	& Monte Carlo		&	& Diode	& Monte Carlo \\
  \hline \hline
  							& \multicolumn{5}{c}{6 MeV} \\
  %d$_\mathrm{max}$ (cm)			& 1.4		& 1.4				&	& 1.4		& 1.4 \\
  d$_{90}$	 (cm)					& 2.0		& 1.9				&	& 2.0		& 1.9 \\
  d$_{50}$	 (cm)					& 2.4		& 2.4				&	& 2.5		& 2.4 \\
  d$_{20}$	 (cm)					& 2.7		& 2.7				&	& 2.7		& 2.7 \\
  FWHM$_\mathrm{cross}$(cm)	& 14.9	& 14.8			&	& 15.1	& 15.0 \\
  FWHM$_\mathrm{in}$ (cm)		& 14.8	& 14.9			&	& 14.9	& 14.9 \\
  80-20\%$_\mathrm{cross}$ (cm)	& 2.2		& 1.9				&	& 2.2		& 1.8	 \\
  80-20\%$_\mathrm{in}$ (cm)		& 2.2		& 1.9				&	& 2.2		& 1.9	 \\
							& \multicolumn{5}{c}{12 MeV} \\
  %d$_\mathrm{max}$ (cm)			& 2.6		& 2.8				&	& 2.6		& 2.8 \\
  d$_{90}$	 (cm)					& 4.0		& 4.0				&	& 3.9		& 3.9 \\
  d$_{50}$	 (cm)					& 5.0		& 5.0				&	& 5.0		& 5.0 \\
  d$_{20}$	 (cm)					& 5.5		& 5.6				&	& 5.5		& 5.6 \\
  FWHM$_\mathrm{cross}$(cm)	& 15.5	& 15.4			&	& 15.5	& 15.4 \\
  FWHM$_\mathrm{in}$ (cm)		& 15.3	& 15.4			&	& 15.3	& 15.4 \\
  80-20\%$_\mathrm{cross}$ (cm)	& 1.6		& 1.5				&	& 1.6		& 1.4	 \\
  80-20\%$_\mathrm{in}$ (cm)		& 1.7		& 1.5				&	& 1.6		& 1.5	 \\
    							& \multicolumn{5}{c}{20 MeV} \\
  %d$_\mathrm{max}$ (cm)			& \bf{2.3}	& \bf{1.9}			&	& \bf{1.8}	& \bf{2.5} \\
  d$_{90}$	 (cm)					& 5.8		& 5.7				&	& 5.7		& 5.7 \\
  d$_{50}$	 (cm)					& 8.4		& 8.3				&	& 8.4		& 8.3 \\
  d$_{20}$	 (cm)					& 9.2		& 9.5				&	& 9.2		& 9.5 \\
  FWHM$_\mathrm{cross}$(cm)	& 15.1	& 15.0			&	& 15.1	& 15.0 \\
  FWHM$_\mathrm{in}$ (cm)		& 15.1	& 15.2			&	& 15.1	& 15.2 \\
  80-20\%$_\mathrm{cross}$ (cm)	& 1.0		& 0.8				&	& 1.0		& 0.8 \\
  80-20\%$_\mathrm{in}$ (cm)		& 1.0		& 0.9				&	& 1.0		& 0.8	 \\
  \hline \hline
  
  \end{tabular}
  \caption{Measured and simulated dose parameters for TrueBeam electron fields at 6, 12 and 20 MeV for $20\times20$ cm$^2$ apertures with jaws set to $21\times21$ and $40\times40$ cm$^2$. Width metrics are given in crossline (cross) and inline (in) orientations.}
  \label{tab:doseChar20}
\end{table}

	Common metrics used to characterize depth-dose and dose profiles (i.e. d$_\mathrm{90}$, d$_\mathrm{50}$, d$_\mathrm{20}$, FWHM and 80-20\% penumbra width) are summarized in Table \ref{tab:doseChar05} for $5\times5$ cm$^2$ apertures with jaws set to $6\times6$ and $40\times40$ cm$^2$, and in Table \ref{tab:doseChar20} for $20\times20$ cm$^2$ apertures with jaws set to $21\times21$ and $40\times40$ cm$^2$. Agreement between Monte Carlo and measurement at 12 and 20 MeV is well within 2 mm for most metrics. At 6 MeV, depth-dose metrics are within 1 mm between Monte Carlo and measurement while profile metrics are mostly within 3 mm. The disagreement observed in profile widths at 6 MeV is due, primarily, to the discrepancy in penumbra slope between measurement and simulation at all field sizes and jaw settings. Note that any differences in measured and simulated FWHM represent twice the distance to agreement.

	Table \ref{tab:doseGamma} summarizes pass-statistics for a one-dimensional gamma comparison of Monte Carlo data against measurement with 2\%/2 mm pass criteria and 5\% cut-off. Pass-statistics for 12 and 20 MeV are excellent and greater than 98\% for all data sets. At 6 MeV, pass-statistics are excellent in depth, while $20\times20$ cm$^2$ profiles agree better than $5\times5$ cm$^2$ profiles. This is not surprising since the penumbra makes up a large portion of a $5\times5$ cm$^2$ field and Figures \ref{fig:jawInMC} and \ref{fig:jawOutMC} show that the 6 MeV model performs worst in the penumbra. When pass criteria are extended to 3\%/3 mm, most of the 6 MeV data sets have 100\% pass rates, and the remainder have pass rates higher than 80\%.

\begin{table}[h]
  \centering
  \begin{tabular}{ c  c  c  c  c  c  c  c  c  }
  \hline
  Field Size 				& \multirow{2}{*}{Metric}		& \multirow{2}{*}{6 MeV}	& \multirow{2}{*}{12 MeV}	& \multirow{2}{*}{20 MeV}	&	& \multirow{2}{*}{6 MeV}	& \multirow{2}{*}{12 MeV}	& \multirow{2}{*}{20 MeV} \\  
  (cm$^2$)				&						&					&					& 					&	&					&					&\\
  \hline \hline
  						&						& \multicolumn{3}{c}{$21\times21$ cm$^2$}							&	& \multicolumn{3}{c}{$40\times40$ cm$^2$} \\
  \multirow{5}{*}{$20\times20$}	& depth					&100					& 100				& 99					&	& 100				& 100				& 100 \\
						& crossline$_\mathrm{max}$	& 88					& 100				& 100				&	& 87					& 100				& 99 \\
						& inline$_\mathrm{max}$		& 94					& 100				& 100 				&	& 92					& 100				& 99 \\
						& crossline$_\mathrm{50}$	& 93					& 100				& 100 				&	& 91					& 100				& 100 \\
						& inline$_\mathrm{50}$		& 99					& 100				& 100 				&	& 96					& 100				& 100 \\
						&						&					&					& 					&	&					&					&\\
						&						& \multicolumn{3}{c}{$6\times6$ cm$^2$}								&	& \multicolumn{3}{c}{$40\times40$ cm$^2$} \\
  \multirow{5}{*}{$5\times5$}	& depth					& 100				& 100				& 99	 				&	& 100				& 100				& 99 \\
						& crossline$_\mathrm{max}$	& 67					& 100				& 100 				&	& 66					& 100				& 98 \\
						& inline$_\mathrm{max}$		& 78					& 100				& 100 				&	& 70					& 100				& 98 \\
						& crossline$_\mathrm{50}$	& 73					& 100				& 100 				&	& 73					& 100				& 100 \\
						& inline$_\mathrm{50}$		& 100				& 100				& 100 				&	& 78					& 100				& 100 \\
  \hline \hline
  
  \end{tabular}
  \caption{One-dimensional gamma-pass statistics for simulated depth and profile curves compared against measured data with the indicated jaws settings. Pass statistics are given as a percent of dose points that pass within 2\%/2 mm with a 5\% cut-off.}
  \label{tab:doseGamma}
\end{table}

\subsubsection{Jaw-dependent energy fluence}

\begin{figure}[h]
  \caption{Charged particle energy fluence through a $5\times5$ cm$^2$ MLC aperture with $6\times6$ and $40\times40$ cm$^2$ jaw settings, scored above the phantom and plotted in the crossline direction. Profiles represent the charged particle energy fluence normalized by the number of incident electrons simulated. In-field particles refer to those that have not been scattered by the jaws and MLC and are plotted separately from those that have.}
  \centering
  \includegraphics[width=1.0\textwidth]{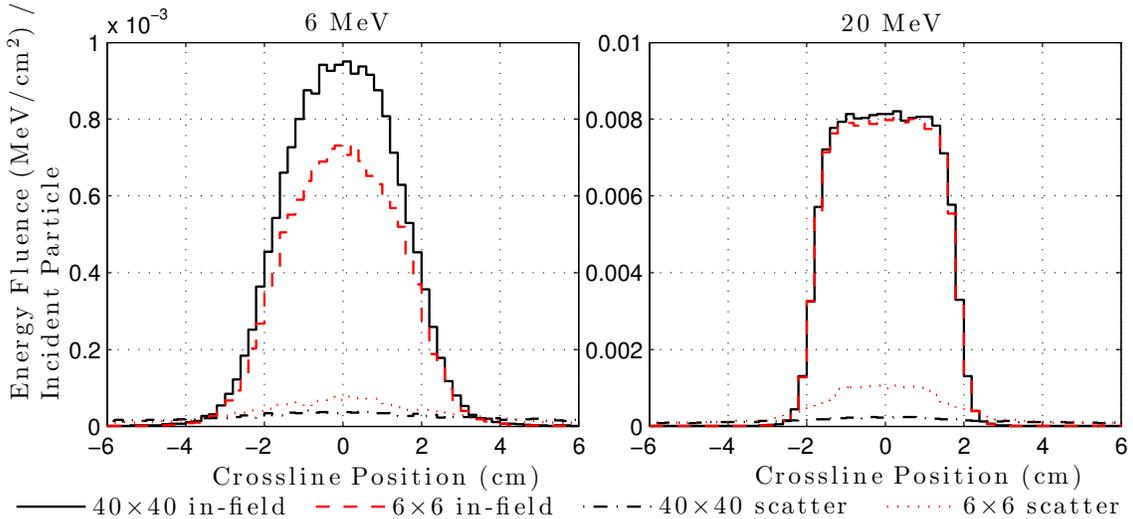}
  \label{fig:jawFluence}
\end{figure}

	Figure \ref{fig:jawFluence} shows simulated energy fluence profiles for charged particles (electrons and positrons) scored above the phantom for a $5\times5$ cm$^2$ MLC aperture. The profiles are plotted in the crossline direction and represent the energy fluence of all of the charged particles passing through a $6\times6$ cm$^2$ cross-section centred on the central axis directly above the phantom, normalized by the number of incident electrons simulated. Data is shown for both $6\times6$ and $40\times40$ cm$^2$ jaw settings at 6 and 20 MeV. The top curves represent charged particle energy fluence that has not interacted with the MLC or jaws (in-field electrons) while the lower curves represent the energy fluence scattered from the jaws or MLC (scattered electrons).
	
	The in-field energy fluence contribution is essentially the same between jaw settings at 20 MeV, while at 6 MeV, there is a 25\% decrease in energy fluence when the jaws are moved in from $40\times40$ to $6\times6$ cm$^2$. At 20 MeV, the dominant feature is a four-fold increase in scatter from the jaws and MLC when the the jaws are moved in from $40\times40$ to $6\times6$ cm$^2$.

%%% 		DISCUSSION			%%%

\section{Discussion}

	Our complete Monte Carlo electron source models agree with measurement within 2\%/2 mm at 12 and 20 MeV, and within 3\%/3 mm for the most part at 6 MeV, which is comparable to the performances of complete accelerator models presented by Salguero et al., Klein et al., and Al-Yahya et al.  \cite{Salguero:2009, Klein:2008, AlYahya:2005mp}. It is well established that electron dose distributions are affected by jaw position, and the data presented show that our models reflect this effect.

	Currently, there is some disagreement as to the most appropriate jaw settings for MLC-shaped MERT fields. From Figure \ref{fig:jawMeas}, it is apparent that electron transmission through the MLCs is negligible and that jaws are unnecessary for shielding in MERT applications. The work by Henzen et al. \cite{Henzen:2014} used a fixed jaw setting of $15\times35$ cm$^2$ for all aperture sizes to simplify calculations. Photon fields, however, generally utilize jaws set just beyond the MLC aperture, so to avoid inefficiencies in MPERT delivery, it would be ideal to use similar jaw settings for both photon and electron fields. Klein et al. \cite{Klein:2008} recommended that jaws be set 1 cm beyond the MLC field periphery except in the case of fields smaller than $6\times6$ cm$^2$ in order to minimize the field penumbra without negatively impacting in-field uniformity. In contrast, studies by Eldib et al. \cite{Eldib:2014} and by Connell and Seuntjens \cite{Connell:2014} into removing or modifying the electron scattering foil for MERT applications suggested that the reduction of bremsstrahlung photon contamination is more desirable than field flatness. In the study by Connell et al., jaws were fixed at $22\times22$ cm$^2$ for all fields and energies. A subsequent paper by Connell et al. \cite{Connell:2014-2} showed the output of small FLEC-shaped electron field to change more than 5\% with jaw position changes as small as 0.5 mm, so, although there is no consensus on the most appropriate jaw settings for MERT applications, it is evident that the dosimetric impact of the jaws must be accurately accounted for.

	In Figure \ref{fig:jawMeas}, it can be seen that jaw position has the greatest impact at small field size ($5\times5$ cm$^2$). By assessing trends in charged particle energy fluence in Figure \ref{fig:jawFluence}, one may observe that the differences in dose at 6 MeV are dominated by the changes in the in-field fluence as a function of jaw position, while dosimetric changes at 20 MeV are dominated by jaw and MLC scatter contributions to the overall energy fluence. Our complete accelerator model is able to simulate these jaw-position dependancies in small fields, and as small fields are likely to be well used in MERT and MPERT, this competency is an asset.

	The good agreement demonstrated in field width at both d$_\mathrm{max}$ and d$_{50}$ illustrates the accuracy of our model with depth at both representative field sizes and for all three energies modelled. The discrepancies in penumbra slope between measurement and simulation at 6 MeV may be due to assumptions made about the mono-energetic or forward-directed nature of the incident electron source, inaccuracies in our electron foil model, or limitations in our MLC model. Geometric inaccuracies in the model are likely to have the greatest impact at low electron energy where the degree of scatter is the greatest.

%%% 		Conclusion		%%%

\section{Conclusion}

	We have modified a complete Monte Carlo model of a Clinac 21EX accelerator to simulate TrueBeam electron sources at 6, 12 and 20 MeV. Simulated data agrees with measured depth and profile data within 2\%/2 mm for the 12 and 20 MeV models and, for the most part, within 3\%/3 mm for the 6 MeV model. Our complete accelerator model allows for simulation of jaw position dependancies and is, therefore, flexible for use in MERT and MPERT planning.

\section*{Acknowledgments}

This work was supported, in part, by a Varian Research Grant.

%\end{spacing}

\section*{References}
\bibliographystyle{ieeetr}
\bibliography{slloyd_paper1draft_arXiv}

\end{document}